%% file: main.tex
\documentclass[10pt,twocolumn]{article}

\usepackage{latex8}
\usepackage{times}
\usepackage{inconsolata}
\usepackage{pifont}
\usepackage{booktabs}
\usepackage{multirow}
\usepackage{tikz}
\usepackage{amsmath,bm,times}
\usepackage{verbatim}
\usepackage{subcaption}
\usepackage{bibspacing}
\usepackage{cite}
\usepackage{url}

\usepackage{fancyhdr}

\fancypagestyle{doi}{\lfoot{$^*$: Corresponding Author: Xiang Chen. 
$^{\sharp}$: Equally contributing authors.
}}

\usepackage[small,compact]{titlesec}

\newcommand{\cx}[1]{\textcolor{black}{#1}}

\begin{document}

\title{\vspace*{-.6in}DeepSCC: Source Code Classification Based on Fine-Tuned RoBERTa\vspace*{-.1in}}

\author{Guang Yang            \and
        Yanlin Zhou           \and
		Chi Yu                \and
		Xiang Chen$^{*}$ 
      }

\affiliation{
      School of Information Science and Technology, Nantong University, Nantong, China\\
      1930320014@stmail.ntu.edu.cn 1159615215@qq.com yc\_struggle@163.com xchencs@ntu.edu.cn
}

\maketitle

\begingroup
\renewcommand{\thefootnote}{}
\footnotetext[1]{* Xiang Chen is the corresponding author.}
\footnotetext[1]{DOI reference number: 10.18293/SEKE2021-005}
\endgroup

\begin{abstract}
In software engineering-related tasks (such as programming language tag prediction based on code snippets from Stack Overflow), the programming language classification for code snippets is a common task. 
In this study, we propose a novel method DeepSCC, which  uses a fine-tuned RoBERTa model to classify the programming language type of the source code. 
In our empirical study, we choose a corpus collected from Stack Overflow, which contains 224,445 pairs of code snippets and corresponding language types. After comparing nine state-of-the-art baselines from the fields of source code classification and neural text classification in terms of four performance measures (i.e., Accuracy, Precision, Recall, and F1), we show the competitiveness of our proposed method DeepSCC.

\end{abstract}

\input{1.introduction}

\input{2.related_work}
\input{3.approach}

\input{4.experiment}
\input{6.conclusion}

\setlength{\bibitemsep}{.075in}
{\footnotesize
  \bibliographystyle{IEEEtran}
\bibliography{references}}
\end{document}

%% file: 1.introduction.tex
\section{Introduction}

Recently, multiple programming languages (such as Java, Python, C++) are often used together in the large-scale software development process, since different development tasks often use different programming languages.
When developers ask questions on Stack Overflow~\cite{chen2019sethesaurus}\cite{cao2021automated}, the answers to the questions are closely related to the type of programming language. Therefore, Stack Overflow needs to use the correct programming language tag of posts to match the related answers for users, and the source code classification task can effectively solve this problem.

% On the other hand, we find that researchers train corresponding models according to different programming languages in the field of automatic code comment generation. For example, according to the characteristics of the Java language, the researchers use the camel case naming convention to segment the identifiers. While according to the characteristics of the Python language, the researchers segment the identifiers according to `\_'. However, in the application, users have to choose the corresponding identifier split method according to the type of programming language. In this scenario, using code classification tools can also effectively solve this type of method selection problem.

In the previous studies, this task is often modeled as a text classification problem. Then machine learning methods can be used to classify the source code's language type. For example, Khasnabish et al.~\cite{khasnabish2014detecting} used a Naive Bayesian classifier. Alrashedy et al.~\cite{alrashedy2020scc++} used a random forest classifier and XGBoost. Motivated by the research progress of neural text classification and code semantic learning~\cite{chen2019deepcpdp}, we propose a novel method DeepSCC by fine-tuning the pre-trained model RoBERTa~\cite{sun2019fine} to perform the source code classification task.

To verify the effectiveness of our proposed method DeepSCC, we choose a corpus collected from Stack Overflow, which contains 224,445 pairs of code snippets and corresponding language types. We first perform data preprocessing on this corpus, such as word segmentation, discarding noisy code snippets. Then, we use the corpus to fine-tuning the RoBERTa model~\cite{sun2019fine}. We compared DeepSCC with nine state-of-the-art baselines. For these chosen baselines, two baselines are selected from the source code classification field~\cite{alreshedy2018scc,alrashedy2020scc++}, and the remaining baselines are selected from the neural text classification field~\cite{DBLP:journals/corr/Kim14f,bojanowski2017enriching,mikolov2013distributed,vaswani2017attention,devlin2018bert}. 
In terms of four performance measures (Accuracy, Precision, Recall, and F1), we find DeepSCC can outperform these baselines.

The main contributions of our study can be summarized as follows:

(1) We propose a novel method DeepSCC by fine-tuning the pre-trained model RoBERTa~\cite{sun2019fine}, which can classify the language type of the source code. We share our trained classification model for other researchers to follow and replicate our study\footnote{\url{https://huggingface.co/NTUYG/DeepSCC-RoBERTa}}.

(2) We choose corpus gathered from Stack Overflow as our experimental subject. Then we choose two baselines proposed by Alrashedy et al.~\cite{alrashedy2020scc++} (i.e., in the source code classification field) and seven baselines based on TextCNN and Transformer (i.e., in the neural text classification field). Final experimental results show that DeepSCC can improve the performance of source code classification.

% \item In addition to comparing with SCC++, we also compare the classification models composed of TextCNN and Transformer in the field of neural text classification. Empirical results show the classification performance of our proposed method DeepSCC is also the best.

%% file: 2.related_work.tex
\section{Related Work}

% \subsection{Source Code Classification}

In previous studies on source code classification, 
Kennedy et al.~\cite{van2016software} proposed a software language model to recognize the entire source code file from Github. Their classifier is based on five natural language  statistical  models. They gathered corpus  from GitHub and considered 19 programming languages. 
Khasnabish et al.~\cite{khasnabish2014detecting} collected more than 20,000 source code files. 
These source codes are downloaded from multiple repositories in GitHub. The model uses the Bayesian classifier and aims to predict ten programming languages.  
Klein et al.~\cite{klein2011algorithmic} collected 41,000 source code files from GitHub as the training set and randomly selected 25 source code files as the test set. However, their methods, which are based on supervised learning and feature selection methods, can only achieve 48\% accuracy at most. 
Alrashedy et al.~\cite{alreshedy2018scc} proposed the method SCC  to classify source code snippets via Naive Bayes classifier, with an accuracy of about 75\%. This method can also distinguish programming language families (such as C, C\# and C++) with an accuracy of 80\%, and can identify programming language versions (such as C\#3.0, C\#4.0, and C\#5.0) with an accuracy of 61\%. 
Recently, Alrashedy et al.~\cite{alrashedy2020scc++} classified the language types for code snippets in  Stack Overflow. They used the random forest classifier and XGBoost to build classifiers. 
%Then they further combined the text information of the questions in Stack Overflow and code snippets to improve accuracy. 
%Gilda~\cite{gilda2017source} used a multi-layer neural network-word embedding layer and convolutional neural network to classify source code. Their constructed model can classify 60 languages. 
Different from the previous studies, we are the first to introduce a pre-trained model to this task and then proposed a novel method DeepSCC. The final results show the competitiveness of our proposed method when compared to state-of-the-art baselines.

%In this paper, we introduce a pre-training model to this task for the first time and proposed a novel method DeepSCC. 

% \subsection{Code Understanding}

% Code understanding is one of the active topics, and many researchers have conducted various studies on source code understanding. In addition to source code classification, researchers are doing code comment generation and nl2code (natural language conversion into code). In the research direction, researchers have made many research achievements. For example, Chen et al. summarized many research work on code comment generation. Yin et al.~\cite{yin2017syntactic} proposed a neural network structure based on a grammatical model, which can parse natural language descriptions into source code written in general programming languages (such as Python).

%% file: 3.approach.tex
\section{Method}

\subsection{Overview of DeepSCC}

\begin{figure*}[htbp]
	\centering
  \vspace{-1mm}
	\includegraphics[width=1\textwidth]{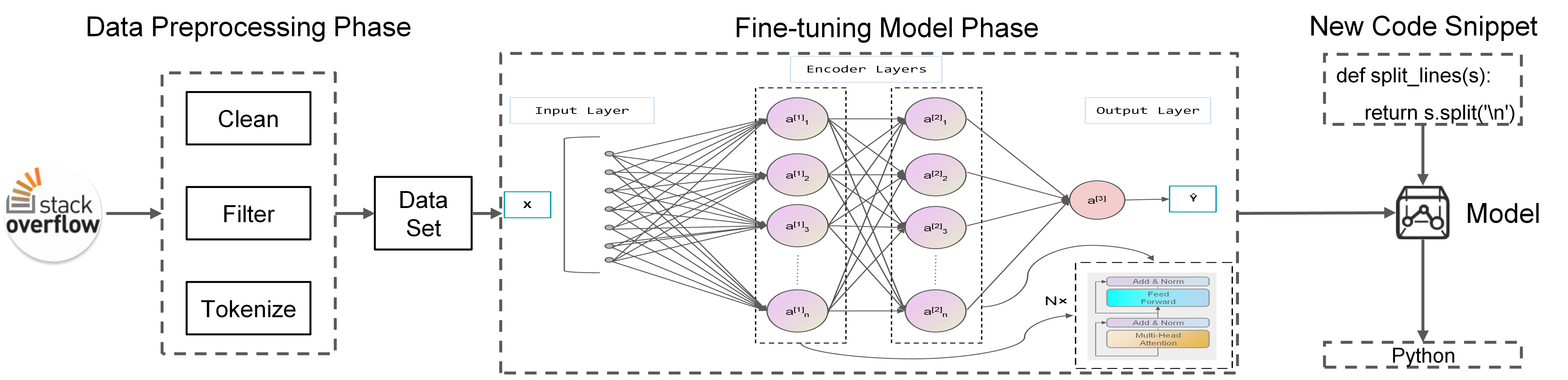}
	\caption{The framework of our proposed method DeepSCC}
  \vspace{-1mm}
	\label{fig:framework}
\end{figure*}

In this section, we show the framework of DeepSCC in Figure~\ref{fig:framework}.
% In our method, we use the fine-tuning RoBERTa model~\cite{sun2019fine} to do downstream code classification task. Specifically, we first preprocess the corpus, including data cleaning, filtering and word segmentation. Then, the processed code fragment sequence is used as the input of the model, and the code semantic learning is performed through the two-layer Transformer encoder to obtain the semantic representation information of the code. Finally, a linear layer and Softmax function are used to predict the type of programming language.
In particular, we first preprocess the corpus, including data cleaning, filtering, and word segmentation. Then we fine-tune the pre-trained model RoBERTa to predict the type of programming language.

\subsection{Data Preprocessing Phase}
In this phase, the data cleaning and filtering are consistent with previous work~\cite{alrashedy2020scc++}.
However, we find that the previous code classification methods treat the code word as the basic unit. Its disadvantage is that it cannot effectively solve the out-of-vocabulary (OOV) problem. That means there exist some words, which are not in the training set but in the testing set.
To solve the OOV problem, we use the Byte-Pair Encoding (BPE) proposed by Sennrich et al.~\cite{sennrich2015neural}. It is a mixture between character-level and word-level representations. Using BPE can avoid a large number of "[UNK]" symbols in the test set, as "[UNK]" symbols may decrease the performance of the pre-trained model. For example, the original code snippet is ``def split\_lines(s): return s.split(`$\backslash$n')", and the result after using BPE segmentation is  ``def", ``Ġsplit", ``\_", ``lines", ``(", ``s", ``)", ``:", ``Ġreturn", ``Ġs", ``.", ``split", ``(", ``'", ``Ċ", ``'", ``)". \cx{Here Ġ means that it is the first subword of a subword division.}

\subsection{Fine-tuning Model Phase}

In this phase, we continue to pre-train the RoBERTa model on the code corpus with the MLM (mask language model) method, and then use the pre-trained model to fine-tune the code classification task. RoBERTa~\cite{sun2019fine} is similar to Bert (Bidirectional Encoder Representations from Transformers)\cite{devlin2018bert}, and DeepSCC uses Transformer as the method's main framework because Transformer can more thoroughly capture the bidirectional relationship in the text. 
In particular, we treat the code as text and use the method MLM  to constantly fine-tuning the roberta-base model\footnote{\url{https://huggingface.co/roberta-base}} on the corpus to obtain our pre-trained language model. During the fine-tuning process, we do not tune the parameters of the model's bias and LayerNorm.weight weights, and use the AdamW method to fine-tune the other parameters.

Consider that different layers of the neural network can capture different levels of syntactic and semantic information. In our study, we choose the last layer of Encoder as the feature representation of the whole code snippet, feed it into the linear layer, and obtain the model prediction label by Softmax, which can be used to calculate the cross-entropy loss with the real label. Then we use AdamW as the optimizer to perform gradient descent and back propagation to update the model parameters. Finally, we can obtain our fine-tuned model.

% Specifically, RoBERTa~\cite{sun2019fine} uses Transformer's Encoder as the main framework of the algorithm for semantic learning. To more thoroughly capture the two-way relationship in the sentence, RoBERTa learns through the Bidirectional Encoder. Given a piece of code, after BPE segmentation, the input of the model $X=(x_{1},x_{2},\cdots,x_{n})$ is obtained. Through the learning of $Position$ $Embedding$, $Segment$ $Embedding$ and $Token$ $Embedding$ layers, the three learned embedding vectors are superimposed to obtain the embedding feature vector $X$ of the code fragment. Then, the embedded feature vector $X$ is input into the bidirectional Encoder for semantic learning. First, the model inputs the vector $X$ into the $multi$-$head$ $attention$ layer. Later, the model uses $residual$ $connection$ and $layer$ $normalization$ to make the matrix operation dimension consistent and normalize the hidden layer in the network to a standard normal distribution, which can accelerate the model training speed and the convergence. In the next step, the model passes the $feed$-$forward$ layer. Then the model also uses $residual$ $connection$ and $layer$ $normalization$ to generate the semantic representation vector $X_{semantic}$. Finally, the semantic vector $X_{semantic}$ is mapped to the programming language category label through a linear layer, and the corresponding programming language is obtained through the Softmax algorithm.

%% file: 4.experiment.tex
\section{Experiment}

\subsection{Experimental Subject}

We choose the corpus shared by Alreshedy et al.~\cite{alreshedy2018scc} as our experimental subject.
%This corpus comes from the Posts.xml\footnote{\url{https://archive.org/download/stackexchange/stackoverflow.com-Posts.7z}} file on StackOverflow. 
Alreshedy et al. gathered code snippets from 21 programming languages (i.e., Bash, C, C\#, C++, CSS, Haskell, HTML, Java, JavaScript, Lua, Objective-C, Perl, PHP, Python, R, Ruby, Scala, SQL, Swift, Visual Basic, and Markdown).
After manual analysis on their gathered corpus, we find: (1) The number of the code snippets related to Markdown is only 1,210, which is significantly lower than that of other languages. (2) In the code snippets related to HTML, we find most of these code snippets  also include CSS and JavaScript code segments. Therefore, we remove the code snippets related to these two languages.
Finally, we use 179,556 code snippets for model training and 44,889 code snippets for model testing via stratified sampling.

\subsection{Performance Measures}

To  compare the performance between our proposed method and the baselines, we choose the following four performance measures: Accuracy, Precision, Recall, and F1. Before introducing these measures, we first illustrate the following concepts: 

\begin{itemize}
\item True Positive (TP): The positive sample is successfully predicted as positive.
\item True Negative (TN): The negative sample is successfully predicted as negative.
\item False Positive (FP): The negative sample is wrongly predicted as positive.
\item False Negative (FN): The positive sample is wrongly predicted as negative.
\end{itemize}

Then the four performance measures can be computed as follows:

\begin{equation}
\text {Accuracy}=\frac{TP+TN}{TP+TN+FP+FN}
\end{equation}
\begin{equation}
\text { Precision }=\frac{T P}{T P+F P} 
\end{equation}
\begin{equation}
\text { Recall }=\frac{T P}{T P+F N} 
\end{equation}
\begin{equation}
\text { F1}=\frac{2 \times \text { Precision } \times \text { Recall }}{\text { Precision }+\text { Recall }}
\end{equation}

Accuracy is the most intuitive performance measure, and it is the ratio of the correctly predicted observations to the total observations.
Precision is the ratio of the correct predicted positive observations to the total predicted positive observations.
Recall indicates how many positive examples in the sample are predicted correctly.
F1 is the average of Precision and Recall.

\subsection{Baselines}

In the RQ, we first compare our proposed method DeepSCC with two state-of-the-art methods from source code classification (i.e., Random Forest and XGBoost methods used in SCC++~\cite{alrashedy2020scc++}). We also choose TextCNN~\cite{DBLP:journals/corr/Kim14f} and Transformer~\cite{vaswani2017attention} with/without pre-trained word vectors (i.e., FastText~\cite{bojanowski2017enriching} or Word2Vec~\cite{mikolov2013distributed}) from the neural text classification field as baselines. Besides, we also select BERT~\cite{devlin2018bert} as a baseline for the pre-trained model.
% In our study, we re-implemented these baselines according to the corresponding description by Pytorch 1.6.0. 

% According to the previous studies' description, these baselines are re-implemented by Pytorch. Finally, we use the fine-tuning pre-trained model RoBERTa~\cite{sun2019fine} as our DeepSCC method.

\subsection{Implementation Details}
In our study, we use Pytorch 1.6.0 to implement our proposed method. 
For baselines in the source code classification field, we run their shared code on our preprocessed corpus.
For baselines in the neural text classification  field, we re-implemented these baselines according to the corresponding description by Pytorch. For BERT and RoBERTa, we pre-train the model in the transformer library. 

% For baseline SCC++, we ran the corpus again according to their open-source code and got the results. For baseline TextCNN and Transformer, we implemented the relevant code. For Bert and RoBERTa, we pre-train the model in the transformer library and get the final result.

It needs to be noticed that pre-trained models (i.e., BERT and RoBERTa) use the method BPE for code segmentation by default. For other baselines, we choose the word\_tokenize method provided by the NLTK library for code segmentation.

We run all the experiments on a computer with an Inter(R) Core(TM) i7-9750H 4210 CPU and a GeForce GTX3090 GPU with 24 GB memory. The running OS platform is Windows 10.

\subsection{Result Analysis}

Table~\ref{tab:table_1} shows the comparison results between DeepSCC and baselines. Table~\ref{tab:table_2} shows the detailed results for each language type in terms of three performance measures (Precision, Recall, and F1) and support (i.e., the number of code snippets related to the given programming language in the test set). 
% Finally, we visualize the result via the confusion matrix in Figure~\ref{fig:result}. The confusion matrix describes the relationship between the true attributes of the sample data and the classification prediction result type in the form of a matrix, and is a common method used to evaluate the performance of the classifier. By using the heat map to visualize the confusion matrix, we can observe the relationship between the predicted value and the true value. Each cell codes the accuracy of the prediction result, and the darker the color indicates the better the classification performance of the category.

\begin{table}[]
 \caption{The comparison results between DeepSCC and baselines}
  \vspace{-2mm}
\resizebox{0.5\textwidth}{!}{
\begin{tabular}{ccccc}
\hline
Method               & Accuracy(\%) & Precision(\%) & Recall(\%) & F1(\%) \\ \hline
Random Forest         & 78.728       & 79.362        & 78.825     & 78.874       \\
XGBoost              & 78.803       & 79.925        & 78.891     & 79.217       \\ \hline
TextCNN              & 82.662       & 83.561        & 82.706     & 82.964       \\
TextCNN+FastText     & 84.201       & 84.719        & 84.285     & 84.369       \\
TextCNN+Word2Vec     & 84.600       & 85.071        & 84.677     & 84.764       \\
Transformer         & 79.035       & 79.801        & 79.107     & 79.272       \\
Transformer+FastText & 75.624       & 76.026        & 75.526     & 75.986       \\
Transformer+Word2Vec & 74.325       & 75.050        & 73.243     & 73.765       \\ 
BERT                & 86.946       & 87.292        & 87.004     & 87.116       \\ \hline
DeepSCC              & \textbf{87.202}       & \textbf{87.424}        & \textbf{87.276}     & \textbf{87.135}       \\ \hline
\end{tabular}
}
 \vspace{-2mm}
\label{tab:table_1}
\end{table}

\begin{table}[]
 \vspace{-2mm}
 \caption{The detailed performance for each programming language}
\begin{tabular}{ccccc}
\hline
            & Precision & Recall & F1 & Support \\ \hline
Bash        & 0.89      & 0.84   & 0.87     & 2427    \\
C           & 0.79      & 0.84   & 0.81     & 2396    \\
C\#         & 0.82      & 0.83   & 0.83     & 2407    \\
C++         & 0.82      & 0.82   & 0.82     & 2442    \\
CSS         & 0.85      & 0.89   & 0.87     & 2362    \\
Haskell     & 0.91      & 0.94   & 0.93     & 2320    \\
Java        & 0.85      & 0.87   & 0.86     & 2417    \\
JavaScript  & 0.83      & 0.82   & 0.82     & 2459    \\
Lua         & 0.92      & 0.90   & 0.91     & 1647    \\
Objective-C & 0.90      & 0.94   & 0.92     & 2410    \\
Perl        & 0.87      & 0.85   & 0.86     & 2378    \\
PHP         & 0.81      & 0.86   & 0.83     & 2455    \\
Python      & 0.85      & 0.87   & 0.86     & 2445    \\
R           & 0.92      & 0.93   & 0.92     & 2362    \\
Ruby        & 0.90      & 0.85   & 0.88     & 2390    \\
Scala       & 0.94      & 0.92   & 0.93     & 2341    \\
SQL         & 0.86      & 0.84   & 0.85     & 2410    \\
Swift       & 0.96      & 0.92   & 0.94     & 2474    \\
VB.Net      & 0.92      & 0.85   & 0.88     & 2347    \\ \hline
\end{tabular}
 \vspace{-2mm}
\label{tab:table_2}
\end{table}

% \begin{figure}[htbp]
% 	\centering
%   \vspace{-1mm}
% 	\includegraphics[width=0.5\textwidth]{fig/Confusion.png}
% 	\caption{The confusion matrix visualization via the heat map}
%   \vspace{-1mm}
% 	\label{fig:result}
% \end{figure}

According to the analysis of the experimental results, we can find:
(1) From Table~\ref{tab:table_1}, we can find that our method can outperform baselines and achieves the best performance in source code classification. Specifically,
it can achieve a maximum performance improvement of 17\%, 16\%, 19\%, and 18\% in terms of Accuracy, Precision, Recall, and F1 respectively. 
The results show that the two-way transformer encoder can learn the deep semantics of the code snippets more effectively, which is helpful to obtain a better classification performance.
% (2) For baselines in the field of source code classification, the use of traditional machine learning model classification has no advantage when compared to deep learning-based methods.
(2) Not all the baselines in the neural text classification field outperform the baselines in the code classification field. That means some traditional machine learning methods can outperform deep learning-based methods in this task.
(3) For baselines in the field of neural text classification, Transformer is not as effective as TextCNN in code classification. This may be because Transformer learns too little code semantics. After adding pre-trained word vectors (such as Word2Vec and FastText), the performance of TextCNN can be slightly improved, but the performance of Transformer is decreased. This shows that pre-trained word vectors can better capture the feature representation of the code when the structure is CNN in this task.
(4) From Table~\ref{tab:table_2}, we can find that DeepSCC can achieve high performance in most of the programming languages. Then we  analyze the cause of the poor performance when the programming languages are C/C++ and CSS/JavaScript. Specifically, 8\% of the code snippets with the actual category of C are predicted to be C++, and 10\% of the code snippets with the actual category of C++ are predicted to be C. Since C++ is almost a superset of C, this indicates that some  C++ code snippets and C code snippets  are indistinguishable, which poses a challenge for the source code classification problem. 6\% of the code fragments with the actual category of CSS are predicted to be JavaScript, and 7\% of the code snippets with the actual category of JavaScript are predicted to be CSS. Because CSS as a style language often appears in the scripting language JavaScript, it is used to dynamically update page elements. This leads to the simultaneous appearance of JavaScript and CSS in the code snippets, which also poses another challenge for the source code classification problem.

% \noindent\textbf{Conclusion.} In the task of source code classification, our method DeepSCC achieves the best performance. Specifically, DeepSCC can achieve a maximum performance improvement of 17\%, 16\%, 19\%, and 18\% in terms of Accuracy, Precision, Recall, and F1-score respectively. 

%% file: 6.conclusion.tex
\section{Conclusion}

In this paper, we propose a novel method DeepSCC for source code classification, which is based on fine-tuned RoBERTa~\cite{sun2019fine}. To show the effectiveness of DeepSCC, we used four widely used performance measures to evaluate the performance of DeepSCC. The results show the competitiveness of DeepSCC when compared to nine state-of-the-art baselines from the fields of source code classification and neural text classification.

\section*{Acknowledgement}
Guang Yang and Yanlin Zhou have contributed equally to this work and they are co-first authors.
This work is supported in part by 
Natural science research project in Universities of Jiangsu Province (18KJB520041).

% In the future, we want to collect more higher quality corpora and try to further improve the prediction performance of some programming languages (such as C/C++ and CSS/JavaScript). Moreover, we also want to use DeepSCC to support other software engineering tasks (such as the model selection for code comment generation task).